\documentclass[aps,twocolumn,showpacs,preprintnumbers,prb,superscriptaddress]{revtex4-2}
\usepackage{graphicx}  
\usepackage{dcolumn}   
\usepackage{bm}        
\usepackage{amssymb}   
\usepackage{framed}
\usepackage{amsmath}
\pdfoutput=1

\usepackage{hhline}
\usepackage{enumitem}
\usepackage{subfigure,amsmath,verbatim,moreverb}
\usepackage{tabularx}
\usepackage{lipsum}
\usepackage{longtable}
\usepackage{booktabs}
\usepackage{threeparttable}
\usepackage{booktabs,amssymb,siunitx}
\usepackage{latexsym}
\usepackage{dcolumn}
\usepackage{amsmath}
\usepackage{epsf}
\usepackage{float}
\usepackage{hyperref}
\usepackage{xcolor}
\usepackage[utf8]{inputenc}
\usepackage[normalem]{ulem}
\usepackage{amsmath}
\usepackage{etoolbox}
\usepackage{soul}
\newcounter{subeqn} %

\usepackage{tikz}
\usepackage{xcolor}
\usetikzlibrary{
  shapes,
  positioning,
  arrows.meta,
  decorations.pathreplacing, 
  calc
}
\usepackage[utf8]{inputenc}
\usepackage[T1]{fontenc}
\usepackage{mathptmx}

\usepackage{booktabs}
\usepackage{tabularx}
\usepackage{adjustbox}
\usepackage{rotating}
\usepackage{color,soul}
\usepackage{multirow}
\usepackage{relsize}
\usepackage{ragged2e}

\usepackage{hyperref}
\hypersetup{
    unicode=true,          
    pdftoolbar=true,        
    pdfmenubar=true,        
    pdffitwindow=false,     
    pdfstartview={FitH},    
    colorlinks=true,       
    linkcolor=blue,          
    citecolor=blue,        
    urlcolor=blue,
}

\usepackage[us,12hr]{datetime} 
\begin{document}

\title{
Electric-Field and Doping-Induced Non collinear Magnetic Interactions in Monolayer Ti$_2$Si
}

\author{Dimple Rani}
\altaffiliation{Corresponding author: dimple.rani@niser.ac.in}
\affiliation{School of Physical Sciences, National Institute of Science Education and Research, An OCC of Homi Bhabha National Institute, Jatni 752050, India}

\author{Gayatri Panda}
\affiliation{School of Physical Sciences, National Institute of Science Education and Research, An OCC of Homi Bhabha National Institute, Jatni 752050, India}

\author{Subrata Jana}
\affiliation{Institute of Physics, Faculty of Physics, Astronomy and Informatics, Nicolaus Copernicus University in Toru\'n, ul. Grudzi\k{a}dzka 5, 87-100 Toru\'n, Poland}

\author{Prasanjit Samal}
\affiliation{School of Physical Sciences, National Institute of Science Education and Research, An OCC of Homi Bhabha National Institute, Jatni 752050, India}

\begin{abstract}

Two-dimensional (2D) silicides are an emerging class of materials whose
magnetic and relativistic properties remain largely unexplored. Using
first-principles calculations, we investigate how electric-field
modulation and transition-metal doping influence the magnetic exchange,
magnetocrystalline anisotropy, and antisymmetric
Dzyaloshinskii–Moriya interaction (DMI) in monolayer Ti$_2$Si. Pristine
Ti$_2$Si is a dynamically stable ferromagnetic metal with in-plane
anisotropy and centrosymmetric bonding, which suppresses DMI even under
strong perpendicular electric fields. To overcome this symmetry
constraint, we introduce Pt and Co substitution at Ti sites. Co enhances
the magnetic exchange, whereas Pt provides strong spin orbit coupling (SOC),
and the combined chemical asymmetry breaks inversion symmetry
sufficiently to induce a sizable DMI. A Wannier-based tight-binding
model captures the orbital-resolved superexchange pathways and reveals a
clear hierarchy between a weak Si-mediated channel and a dominant
Pt-mediated interlayer channel. First Principle calculations confirm that the
Pt-assisted pathway governs the magnitude and sign of the total DMI.
Among all configurations, Pt$_{0.5}$CoTi$_{0.5}$Si exhibits the strongest chiral
interaction, with its intralayer and interlayer contributions favoring opposite
rotation senses, namely counterclockwise (CCW) and clockwise (CW). Our results establish chemically engineered Ti$_2$Si monolayers as a promising
platform for realizing and tuning chiral magnetic textures in 2D silicides.

\end{abstract}

\maketitle

\section{Introduction}

Magnetic textures arise from the delicate balance between symmetric exchange interactions~\cite{bogdanov1989thermodynamically}, magnetic anisotropy~\cite{WANG1996337}, and the antisymmetric Dzyaloshinskii–Moriya interaction (DMI)~\cite{heide2008dzyaloshinskii,birss1964symmetry}. The DMI stems from strong spin–orbit coupling (SOC) in systems lacking inversion symmetry and promotes chiral spin arrangements, giving rise to spin structures with nontrivial topology. In recent years, DMI-driven magnetic states have gained considerable attention due to the fundamental insights they offer into chiral magnetism and their promising implications for high-density, energy-efficient spintronic applications~\cite{gong2017discovery,burch2018magnetism,fert2013skyrmions}.

In recent years, two-dimensional (2D) magnetism has revealed a diverse range of materials capable of hosting DMI-driven spin textures~\cite{dd,yu2011near,ruff2015magnetoelectric,neutron}. Beyond the well-studied CrI$_3$~\cite{10.1093/nsr/nwy109}, Fe$_3$GeTe$_2$~\cite{Fe-based}, MnSe$_2$~\cite{nano}, and FeTe$_2$~\cite{jj81-ng8g}, long-range magnetic order has also been reported in layered halides such as CrBr$_3$ and CrCl$_3$~\cite{PhysRevB.106.054426,PhysRevB.98.144411}, transition-metal phosphorus trichalcogenides (MPX$_3$; M = Mn, Fe, Ni; X = S, Se)~\cite{PhysRevB.94.184428}, and metallic systems including monolayer VSe$_2$~\cite{vse2} and NbSe$_2$~\cite{nbse2}. Transition-metal dichalcogenides (TMDs), in particular, provide a versatile platform for engineering chiral magnetism, where heavy-element substitution~\cite{8vvn-k9p3,voltage}, Janus engineering [e.g., Cr$_2$X$_3$Y$_3$ (X, Y = Cl, Br, I; X $\neq$ Y)~\cite{jan}, MnXY (X, Y = S, Se, Te; X $\neq$ Y)~\cite{janus}, and CrXTe (X = S, Se)~\cite{crjan}], or interfacing with high-SOC substrates such as Pt, W, and Ir~\cite{moreau2016additive,heinze2011spontaneous,emori2013current} can significantly enhance the DMI.

2D silicides remain largely unexplored in the context of 
chiral magnetism and the DMI. While bulk 
and thin-film silicides have been widely studied for their electronic, 
catalytic, and thermoelectric properties~\cite{cao2015transition,IHOUMOUKO2021114304}, 
their magnetic behavior particularly spin orbit-driven effects has received 
comparatively little attention. Reported magnetic silicides such as GdSi$_2$ and 
TbSi primarily exhibit conventional magnetic ordering without evidence of chiral 
textures~\cite{YUN2006e31,tbsi}. Although recent theoretical work shows that heavy-metal incorporation or inversion-symmetry breaking in transition-metal systems can generate sizable DMI~\cite{fert2013skyrmions,jan}, most demonstrations in silicides have been limited to bulk or epitaxial MnSi-based films~\cite{mnsi,mnsi111}.
 Whether 
monolayers or quasi-2D silicides can host DMI-driven chiral magnetism, where 
reduced dimensionality and interface-induced SOC may play a decisive role  
remains an open question. This gap is particularly compelling given the 
structural robustness of silicides, their compatibility with silicon 
technology, and their potential for enhanced SOC.

Titanium silicide (Ti$_2$Si) monolayers have recently been predicted to exhibit remarkable multifunctional properties, including intrinsic ferromagnetism, tunable electronic phases, and catalytic activity~\cite{versa}. First-principles calculations indicate that Ti$_2$Si is dynamically stable and hosts local magnetic moments on Ti atoms arising from partially filled 3$d$ orbitals~\cite{versa}. The magnetic ordering and electronic character can be effectively tuned through strain, carrier doping, or chemical modification, enabling transitions between metallic, half-metallic, and semiconducting states~\cite{lan20212d}. Although Ti and Si possess relatively modest SOC, the Ti 3$d$ states near the Fermi level provide a magnetic framework that can couple to external perturbations, breaking inversion symmetry and enabling the emergence of DMI and chiral spin textures. 

In this work, we investigate monolayer Ti$_2$Si using first-principles
calculations as a 2D transition metal silicide (TMS) platform for tunable magnetism and
potential chiral spin states. We first examine the effect of an external
perpendicular electric field ($\Vec{E}$) on its magnetic anisotropy and exchange
interactions. We then explore substitutional doping with Pt, to enhance
SOC, and Co, to strengthen ferromagnetism, an approach known to
promote sizable DMI in layered systems~\cite{PhysRevLett.115.267210,PhysRevMaterials.6.084401}. 
Together, these two routes allow us to assess both field-induced magnetic
modulation and doping-driven DMI, providing a comprehensive picture of the
conditions under which Ti$_2$Si can sustain noncollinear and chiral magnetic
configurations.

In addition, we construct a minimal Wannier-based tight-binding (TB) model~\cite{pizzi2020wannier90,wan}, which captures the
SOC-induced superexchange pathways responsible for DMI and efficiently verifies the density functional theory (DFT) trends.

\section{Model and Computational Details}
We investigate the 2D TMS compound Ti$_2$Si, which crystallizes in a tetragonal structure belonging to the centrosymmetric space group $P4/mmm$, with optimized lattice constants $a=b=2.73$ $\AA$ with vacumm slab 20 $\AA$. All first-principles calculations are performed within the framework of DFT using the \textsc{VASP} package~\cite{kresse1996efficiency}. 
The electron ion interactions are described using the projector augmented wave (PAW) method~\cite{kresse1996efficient,kresse1999ultrasoft}, 
and the exchange correlation energy is treated within the generalized gradient approximation (GGA) of Perdew, Burke, and Ernzerhof (PBE)~\cite{perdew1996generalized}. 
To account for on-site Coulomb interactions among Ti~$3d$ states, we employ the DFT+$U$ method~\cite{versa} in the Dudarev formulation~\cite{PhysRevB.57.1505} with an effective $U_{\mathrm{eff}} = 2$~eV, 
which correctly reproduces the ground-state magnetic configuration of Ti$_2$Si monolayers. 
The plane-wave basis set is truncated at a kinetic energy cutoff of 420~eV, and Brillouin-zone integrations are performed using a $\Gamma$-centered $16 \times 16 \times 1$ Monkhorst–Pack $k$-point grid. Phonon dispersions are computed using the \textsc{PHONOPY} code~\cite{TOGO20151} within the finite-displacement approach 
using $4 \times 4$ supercells to verify the dynamical stability of all structures.

Using a supercell approach, the DMI vectors ($\vec{d_{ij}}$) are evaluated through a three step procedure. First, full structural relaxations are performed to obtain the equilibrium interfacial geometry, with convergence thresholds of $10^{-6}$~eV in total energy and 0.001~eV/\AA\ in forces. Second, the ground-state charge density is obtained by solving the Kohn--Sham equations without SOC. In the final step, SOC is introduced, and the self-consistent total energy is computed for different orientations of the magnetic moments using VASP’s constrained-magnetization method. This approach, previously employed to assess DMI in bulk frustrated and chiral-lattice magnets~\cite{PhysRevB.84.224429}, is adapted here for the monolayer system. The DMI is extracted by enforcing opposite spin-rotation chiralities (clockwise and counterclockwise) in a $4\times1$ supercell for pristine Ti$_2$Si and a $4\times2$ supercell for Pt- and Co-doped structures. The resulting energy differences yield both the direction and magnitude of the DMI vectors with high accuracy.

To compute the orbital-resolved DMI, we construct a TB Hamiltonian from maximally localised Wannier functions (MLWFs) generated using the \textsc{Wannier90} package~\cite{pizzi2020wannier90}. Our starting point follows the Anderson superexchange framework~\cite{anderson}, later generalized by Moriya~\cite{dmi2} and Dzyaloshinskii~\cite{dmi} to include antisymmetric exchange in the presence of SOC. The Wannier-based TB Hamiltonian~\cite{sutton1988tight} reproduces the \emph{ab initio} electronic structure and provides access to orbital resolved hopping parameters calculated using TBMODELS~\cite{tb-models}, which form the basis for evaluating symmetric and antisymmetric exchange interactions. In this formulation, the MLWFs serve as localized orbitals on both the magnetic cations and the ligand sites, naturally incorporating the underlying crystal symmetry and enabling a rigorous definition of orbital overlaps. SOC is explicitly included in the effective Hamiltonian, and interactions among all relevant nearest neighbours are considered. By analysing the complex hopping amplitudes between orbital pairs, the orbital-resolved DMI is obtained directly for Ti$_2$Si and its doped derivatives.

\section{Results and Discussions}
\subsection{Structural, Electronic, and Magnetic Properties}
\begin{figure*}
    \centering
    \includegraphics[width=1.03\linewidth]{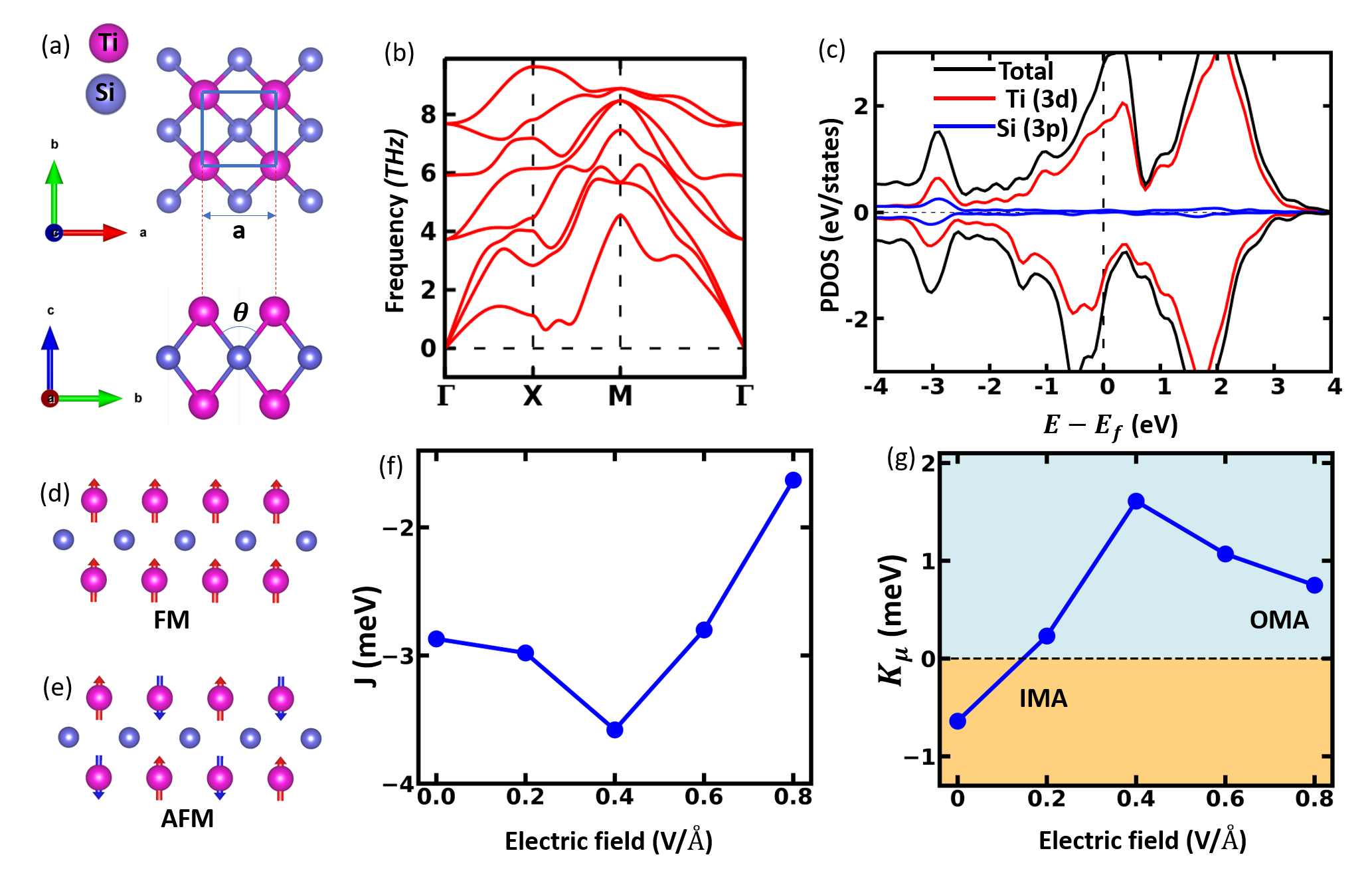}
    \caption{(a) Top and side views of the Ti$_2$Si monolayer, where $a$ denotes the lattice constant and $\theta$ represents the bond angle. (b) Phonon dispersion confirming the dynamical stability of the system. (c) Spin-polarized projected density of states (PDOS) of the Ti$_2$Si monolayer. (d) and (e) Schematic representations of the ferromagnetic and antiferromagnetic spin configurations, respectively. (f) Heisenberg exchange interaction parameter $J$ as a function of external electric field. (g) Single-ion anisotropy as a function of external electric field, where the blue region indicates out-of-plane magnetic anisotropy (OMA) and the yellow region corresponds to in-plane magnetic anisotropy (IMA).}
    \label{Intro}
\end{figure*}

In this paper, we have investigated the monolayer Ti$_2$Si, shown in Fig.~\ref{Intro}(a), which consists of a Ti–Si–Ti trilayer with a bond angle $\theta = 63^{\circ}$, indicating a slight deviation from ideal planarity. The phonon dispersion calculated using a $4\times4$ supercell [Fig.~\ref{Intro}(b)] displays no imaginary frequencies throughout the Brillouin zone, confirming the dynamical stability of the Ti$_2$Si monolayer.
The orbital-resolved DOS in Fig.~\ref{Intro}(c) shows that the states near the Fermi level are mainly contributed by the Ti~3$d$ orbitals, with noticeable hybridization from the Si~3$p$ states. The exchange splitting of the Ti~3$d$ orbitals produces a spin-polarized electronic structure, confirming that Ti$_2$Si behaves as an itinerant ferromagnetic metal. The total magnetic moment is about $1.43~\mu_B$ per unit cell, concentrated mostly on the Ti atoms, with the Si $p$ orbitals contributing through an indirect superexchange pathway. To understand how Ti$_2$Si responds to external perturbations, we applied a perpendicular electric field ($\Vec{E}$) and examined how it affects the electronic and magnetic properties. 

In order to understand the magnetic interactions in Ti$_2$Si monolayers, we consider the generalized spin Hamiltonian,  

\begin{equation}
H = \frac{1}{2}\sum_{i,j} J_{ij} \, \mathbf{S}_i \cdot \mathbf{S}_j 
    + K_{\mu} \sum_i \left(S_i^z\right)^2  
    + \frac{1}{2} \sum_{i,j} \mathbf{d}_{ij} \cdot \left(\mathbf{S}_i \times \mathbf{S}_j\right)~,
    \label{Hamiltomnian}
\end{equation}
where $J_{ij}$ represents the isotropic Heisenberg exchange coupling, $K_{\mu}$ denotes the uniaxial single-ion anisotropy constant, and $\mathbf{d}_{ij}$ is the Dzyaloshinskii--Moriya vector. The anisotropy term governs the preferred spin orientation, stabilizing out-of-plane alignment for $K_{\mu} > 0$ and in-plane orientation for $K_{\mu} < 0$. The DMI contribution encodes the antisymmetric exchange that favors chiral spin textures.
To determine the preferred magnetic configuration, both ferromagnetic (FM) and antiferromagnetic (AFM) alignments were considered [Figs.~\ref{Intro}(d)–(e)]. The Heisenberg exchange interaction parameter $J$ was obtained from the total energy difference of FM and AFM (derived in Supplementary material Section II ~\cite{SM}) 
 as shown in Fig.~\ref{Intro}(f), $J$ remains negative for electric fields in the range $0$–$0.8$~V/\AA, indicating the energetic preference for the FM state. The magnitude of $|J|$ exhibits a nonmonotonic variation with $\Vec{E}$ initially decreasing and then increasing reflecting field-induced modulation of the Ti–$d$ and Si–$p$ orbital overlap that governs the superexchange interaction. The strongest FM coupling is observed near $0.4$~V/\AA, beyond which the interaction weakens.

The single-ion magnetic anisotropy constant ($K_{\mu}$) is evaluated as
\begin{equation}
K_{\mu} = E_{\parallel} - E_{\perp},
\end{equation}
where $E_{\parallel}$ and $E_{\perp}$ denote the total energies with magnetization oriented in-plane and out-of-plane, respectively.  
A positive $K_{\mu}$ indicates an easy-axis anisotropy along the out-of-plane direction, which is favorable for skyrmion formation.  
As shown in Fig.~\ref{Intro}(g), $K_{\mu}$ exhibits strong field dependence: at zero electric field, $K_{\mu}<0$ corresponds to an in-plane magnetic anisotropy (IMA), while increasing the field drives a sign reversal, marking a transition to out-of-plane magnetic anisotropy (OMA).  
This reorientation originates from an electric field induced redistribution of Ti~3$d$ orbital occupations, which alters the SOC contribution to the total anisotropy energy. 
Such  $\Vec{E}$ controlled modulation of $K_{\mu}$ highlights the potential of Ti$_2$Si as a platform for tunable spintronic phenomena.

\begin{figure}
    \centering
    \includegraphics[width=\linewidth]{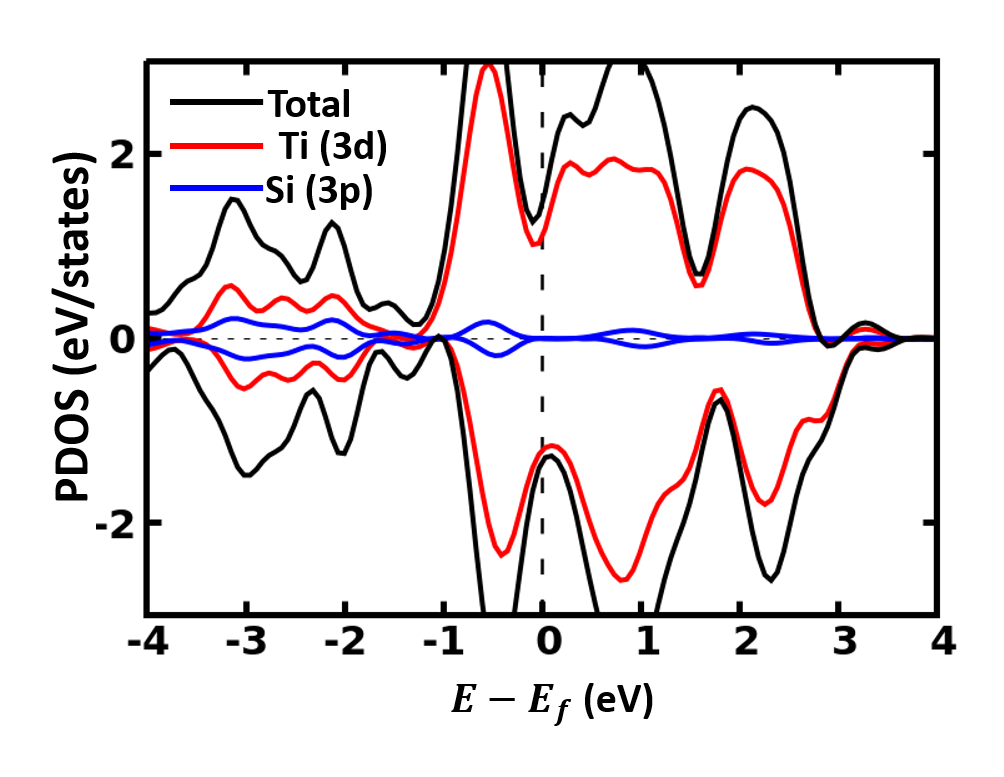}
    \caption{Spin-resolved density of states of the Ti$_2$Si monolayer calculated under an applied electric field of 0.4~V/\AA.}
\label{0.4-fig}
\end{figure}

At $E = 0.4$~V/\AA, the PDOS as shown in Fig.~\ref{0.4-fig}, reveals a clear field-induced redistribution of the
Ti--3d states near the Fermi level, accompanied by changes in Ti--Si
hybridization and local bond geometry. This electronic rearrangement modifies
the relative strength of the competing superexchange channels, leading to a
pronounced minimum in the magnitude of the ferromagnetic exchange interaction
$J$. The magnetic anisotropy $K_\mu$ exhibits a similar nonmonotonic response:
it increases with the applied field and reaches its maximum at
$E = 0.4$~V/\AA, after which it decreases but remains positive. This behavior
reflects the enhanced occupation of Ti--3d states around
$E_F$ at moderate fields, which strengthens the out-of-plane anisotropy before
weakening again at higher fields. Overall, the nonmonotonic evolution of both
$J$ and $K_\mu$ originates from the electric-field-driven reshaping of the Ti--3d
electronic structure.

Overall, the $\Vec{E}$ dependent modulation of both $J$ and $K_{\mu}$ demonstrates that Ti$_2$Si is a promising two-dimensional magnetic system with electrically controllable exchange and anisotropy energies. Such tunability suggests potential applications of Ti$_2$Si in voltage controlled spintronic and magnetoelectric devices.
We have also explored the possibility of inducing inversion symmetry breaking in the Ti$_2$Si monolayer by applying an external electric field, aiming to generate a finite DMI. However, even under a relatively strong electric field of 0.8~V/\AA, the structural inversion symmetry remained preserved, as no significant atomic displacement or potential asymmetry was observed between the two Ti sublayers. This indicates that the intrinsic crystal field and strong covalent bonding in Ti-Si layers effectively screen the external perturbation, thereby suppressing $\Vec{E}$ driven inversion asymmetry and, consequently, the emergence of DMI.

\subsection{Influence of Doping on Exchange interactions and magnetic anisotropy}
\begin{figure*}
    \centering
    \includegraphics[width=1\linewidth]{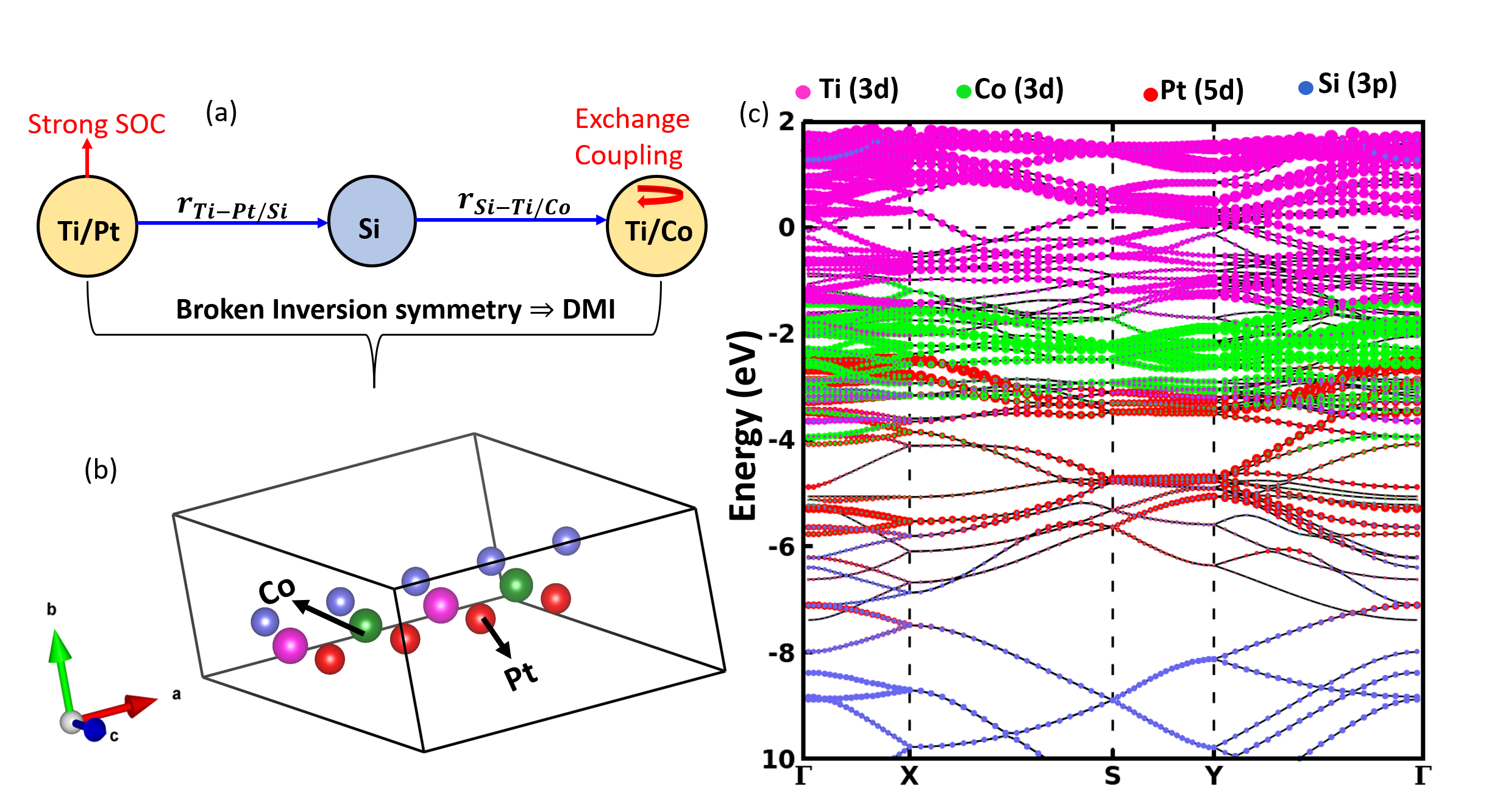}
    \caption{ (a)Schematic representation of the asymmetric Ti/Pt–Si–Ti/Co exchange pathway in the doped Ti$_2$Si monolayer. Pt provides strong SOC on one side and Co strengthens the exchange interaction on the other, together creating the conditions required for a finite DMI. (b) Relaxed atomic structure of the Pt$_{0.5}$CoTi$_{0.5}$Si configuration, illustrating the asymmetric coordination created by simultaneous Pt and Co substitution. 
(c) Orbital-projected band structure of Pt$_{0.5}$CoTi$_{0.5}$Si strcuture.}
    \label{schematic}
\end{figure*}

Even when an external electric field is applied, the inversion symmetry of pristine Ti$_2$Si remains intact, preventing the formation of any finite DMI.  To lift this intrinsic symmetry constraint and strengthen the magnetic interactions, we introduce Transition Metal (TM) substitution  at the Ti sites. Co doping generates substantial local magnetic moments and enhances the exchange coupling, while Pt, owing to its strong SOC, supplies the relativistic contribution required to produce a finite DMI. The schematic Illusration of Co and Pt doping in Ti$_2$Si monolayer is shown in Fig.~\ref{schematic} (a).

To investigate the combined influence of magnetism and SOC, we constructed a $4\times2$ Ti$_2$Si supercell and examined several substitutional configurations, including Pt$_{0.5}$Ti$_{1.5}$Si, Pt$_{0.5}$Co$_{0.5}$TiSi, Pt$_{0.5}$CoTi$_{0.5}$Si, and Pt$_1$Co$_{0.5}$Ti$_{0.5}$Si, together with the limiting cases CoTiSi, PtTiSi, and Pt$_{0.5}$Ti$_{1.5}$Si. Among these, only the Pt$_1$Co$_{0.5}$Ti$_{0.5}$Si configuration is dynamically unstable, while all others remain stable, as confirmed in Fig.~S1 of the Supplementary Material~\cite{SM}. This systematic chemical substitution provides a controlled route to tune the balance between exchange interactions and SOC, offering a microscopic basis for identifying how each contribution influences the emergence and stabilization of chiral magnetic textures. The optimized atomic configuration of Pt$_{0.5}$CoTi$_{0.5}$Si is illustrated in Fig.~\ref{schematic} (b), where Co and Pt atoms are distributed asymmetrically within a $4\times2$ supercell of Ti$_2$Si, thereby breaking local inversion symmetry and providing a natural environment for DMI to emerge.

Fig.~\ref{schematic} (c) shows the orbital-projected band structure of the Pt$_{0.5}$CoTi$_{0.5}$Si configuration, revealing how TM substitution modifies the electronic states of the Ti$_2$Si host. The Pt-derived 5$d$ states strongly hybridize with the Si 3$p$ manifold in the energy window below the Fermi level, reflecting the substantial SOC introduced by Pt. In parallel, the Co 3$d$ orbitals  couple predominantly with the Ti 3$d$ states, forming a broadened $d$-band complex that enhances the exchange-splitting within the magnetic sublattice. The coexistence of these two distinct types of hybridization, Pt-induced strong SOC on one side and Co driven magnetic exchange on the other breaks the local chemical environment and reshapes the low energy band dispersion. This asymmetric redistribution of orbital character provides the microscopic origin for the emergence of finite DMI in the doped Ti$_2$Si system, as it modifies both the relativistic hopping amplitudes and the exchange pathways responsible for chiral magnetic interactions.

The element-resolved magnetic moments listed in Table~\ref{tab1} show a clear dependence on the local Pt/Co/Ti coordination around the magnetic sites. In the Ti-rich composition Pt$_{0.5}$Ti$_{1.5}$Si, the Ti atoms carry a sizeable M$_{\text{Ti}}$ of 1.24~$\mu_B$, while Co is absent and Pt hosts only a very small induced moment of 0.02~$\mu_B$, reflecting the weak polarization of Pt-$5d$ states. Introducing Co into the lattice, as in Pt$_{0.5}$Co$_{0.5}$Ti$_1$Si, leads to the development of a substantial Co moment and a moderate Ti moment, indicating that the spin density becomes redistributed between Co and Ti through hybridization. The induced Pt moment also increases slightly, highlighting the sensitivity of Pt-$5d$ polarization to the presence of magnetic Co neighbors. Upon further increasing the Co concentration to Pt$_{0.5}$Co$_1$Ti$_{0.5}$Si, the Co moment reaches its maximum value (1.55~$\mu_B$), while the Ti moment drops significantly to 0.52~$\mu_B$, demonstrating that the magnetic character becomes increasingly dominated by the Co network. The induced Pt moment is also largest in this composition (0.06~$\mu_B$), consistent with stronger Co–Pt hybridization. In the CoTiSi compound, the Co and Ti moments remain sizable, while in PtTiSi, where Co is absent, Ti retains a moderate moment (0.89~$\mu_B$) and Pt acquires 
only a small induced moment (0.02~$\mu_B$).
Overall, the data reveal a systematic evolution of the magnetic moments: Ti carries the dominant moment in Ti-rich compositions, Co becomes the primary magnetic contributor as its concentration increases, and Pt consistently hosts only a small induced moment whose magnitude reflects the degree of Co–Pt hybridization. These trends highlight the interplay between local chemical coordination and the distribution of spin density across the Pt/Co/Ti sublattice. 

\begin{figure}
    \centering
    \includegraphics[width=0.9\linewidth]{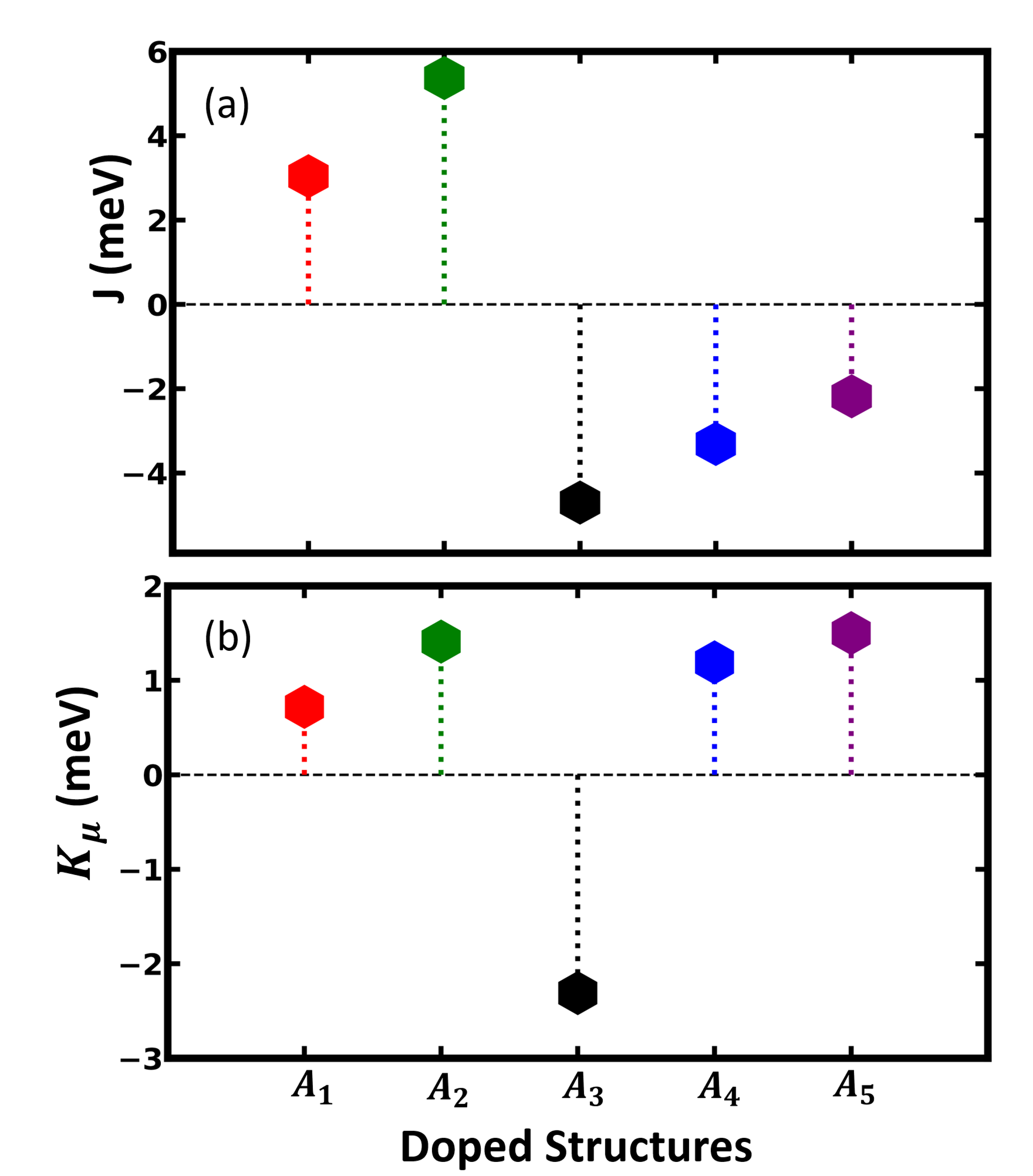}
    \caption{ Evolution of the (a) exchange interaction $J$ and (b) magnetic anisotropy 
$K_{\mu}$ across different doped Ti$_2$Si configurations. The abbreviations 
correspond to: A$_1$ = Pt$_{0.5}$Ti$_{1.5}$Si, A$_2$ = Pt$_{0.5}$Co$_{0.5}$Ti$_{1}$Si, 
A$_3$ = Pt$_{0.5}$Co$_{1}$Ti$_{0.5}$Si, A$_4$ = CoTiSi, and A$_5$ = PtTiSi.
}
    \label{J-K}
\end{figure}

To understand how the local chemical environment shapes the magnetic behavior, we analyzed $J$ and $K_{\mu}$ for the five Pt$_x$Co$_y$Ti$_z$Si compositions listed in Table~\ref{tab1}, with the corresponding trends plotted in Figs.~\ref{J-K} (d)–(e). In the first two compositions, Pt$_{0.5}$Ti$_{1.5}$Si and Pt$_{0.5}$Co$_{0.5}$Ti$_1$Si, both $J$ and $K_{\mu}$ are positive, indicating that the exchange favors AFM alignment and that the magnetic easy axis lies out of the plane.  In these compositions, the combined effect of Ti driven hybridization and SOC naturally favors OMA. When the Co concentration increases further, as in Pt$_{0.5}$Co$_1$Ti$_{0.5}$Si, the magnetic character changes significantly. Here, $J$ becomes strongly negative, signaling the onset of FM coupling driven by enhanced Co-Co superexchange. At the same time, $K_{\mu}$ also turns negative, revealing a rotation of the OMA into IMA, consistent with the more planar bonding geometry around the closely spaced Co atoms. The two end-member compounds, CoTiSi and PtTiSi, likewise show negative $J$, yet retain positive $K_{\mu}$, indicating that although FM exchange dominates, the anisotropy is still governed by the local Ti/Pt coordination and associated SOC. Overall, the trends demonstrate that the sign of $J$ is controlled by the competition between Co-Ti hybridization, which favors AFM, and Co-Co superexchange, which favors FM. In contrast, the sign of $K_{\mu}$ reflects whether the local spin orbit environment stabilizes OMA or IMA. 

\begin{table*}[t]
\centering
\caption{Calculated element-resolved magnetic moments on Ti (M$_{Ti}$) and Co (M$_{Co}$), induced magnetic moment in Pt (M$_{Pt}$) , nearest-neighbor exchange coupling $J$, single-ion anisotropy energy $K_{\mu}$, DMI constants  $d_{1}$ and $d_{1}$, and corresponding Micromagnetic DMI strengths $D_{1}$, $D_{2}$ for the different Pt/Co/Ti/Si compositions considered.}
 \setlength{\tabcolsep}{6.5pt}
\begin{tabular}{cccccccccc}
\hline
\hline
\justifying
Material & M$_{Ti}$ ($\mu_{B}$)&  M$_{Co}$ ($\mu_{B}$)& M$_{Pt}$ ($\mu_{B}$)&$J$ (meV) & $K_{\mu}$ (meV) & $d_{1}$ (meV) &  $d_{2}$ (meV)& D$_1$(mJ/m$^2$)&D$_2$(mJ/m$^2$)\\
\hline
Pt$_{0.5}$Ti$_{1.5}$Si  &  1.24&0.00 &0.02& 3.04  &  0.72  &0.03&-0.12 & 0.05&-0.23 \\
Pt$_{0.5}$Co$_{0.5}$Ti$_{1}$Si &1.17&1.05 &0.05 & 5.37  &  1.41  & 0.09& -0.17 & 0.17&-0.33\\
Pt$_{0.5}$Co$_{1}$Ti$_{0.5}$Si  &0.52&1.55&0.06&-4.70  & -2.31 &0.11 & -0.26 & 0.21&-0.50\\
CoTiSi                        &0.82&0.97 &0.00 & -3.31  &  1.19  & -0.04 &0.00&0.08&0.00\\
PtTiSi &0.89&0.00 &0.02  & -2.18  &  1.50  & 0.00.&-0.02 & 0.00&-0.04\\
\hline
\hline
\end{tabular}
\label{tab1}
\end{table*}

\subsection{Microscopic Origin of DMI: Tight-Binding Perspective}

\begin{figure}
    \centering
    \includegraphics[width=\linewidth]{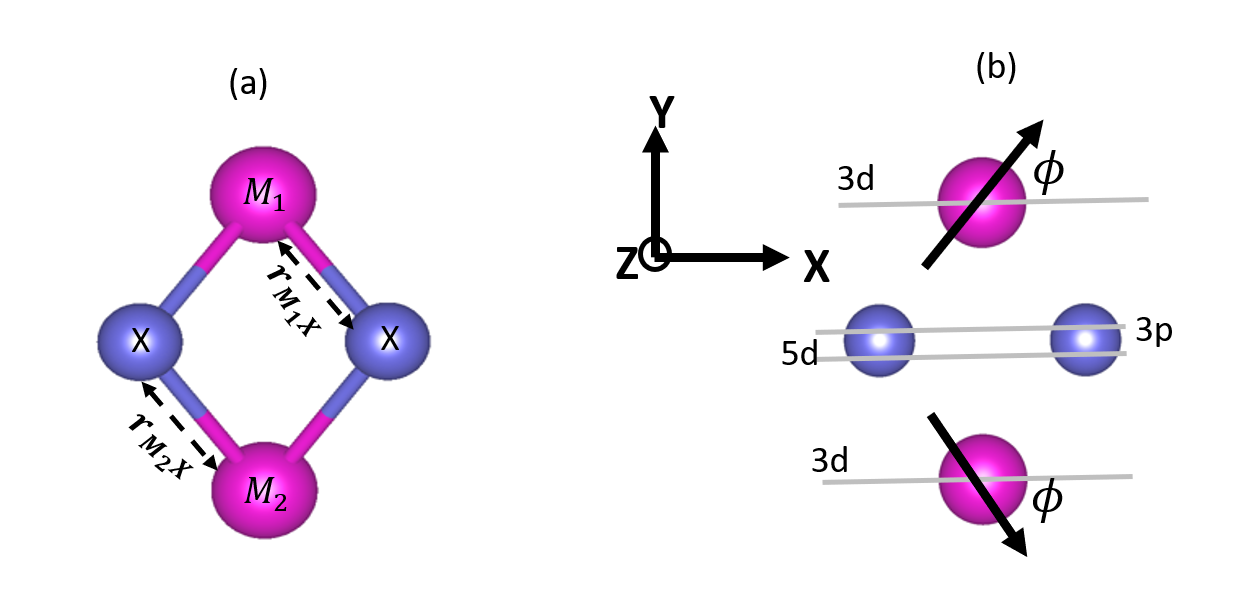}
    \caption{(a) Schematic representation of a TM silicide, where M$_1$ and M$_2$ are magnetic transition-metal sites and X is a
    nonmagnetic ligand. (b) The minimal trimer model consisting of two magnetic
    sites, M$_1$ and M$_2$, each hosting a $d$ orbital, bridged by a nonmagnetic
    ligand X carrying either a $p$ or $d$ orbital. The magnetic moments on
    M$_1$ and M$_2$ are canted by an angle $\phi$ with respect to the $x$-axis.}
    \label{tb1}
\end{figure}

The third term of the Hamiltonian in Eq.~\ref{Hamiltomnian} corresponds to DMI, which arises from SOC in systems without inversion symmetry and plays a crucial role in stabilizing chiral magnetic textures such as skyrmions~\cite{hasan2022first,10.1093/nsr/nwy109}. In the doped Ti$_2$Si lattice considered here, the top and bottom Ti layers are
separated by a nonmagnetic Si spacer, while Pt/Co substitution breaks the horizontal
mirror plane of the pristine Ti$_2$Si structure. Consequently, the system is both
intralayer and interlayer asymmetric~\cite{inter}:\\
(1) Within each magnetic Ti (or Ti/Co) layer, local bond distortions generated by
Pt/Co doping break inversion symmetry, giving rise to intralayer DMI; and\\
(2) The two magnetic layers become chemically and structurally inequivalent,
allowing a distinct interlayer DMI across the Si layer.

To understand the microscopic origin of the resulting noncollinear exchange, we
construct a minimal TB Hamiltonian based on MLWFs~\cite{pizzi2020wannier90}. The model is built on a trimer
M$_1$–X–M$_2$ cluster, where two magnetic sites M$_1$ and M$_2$ are bridged by a
nonmagnetic ligand X, mimicking the Ti–Si–Ti link in the monolayer geometry
[Fig.~\ref{tb1}]. The TB Hamiltonian takes the form
\begin{equation}
\hat{H} = \hat{H}_0 + \hat{H}_t + \hat{H}_{\mathrm{SOC}},
\end{equation}
where $\hat{H}_0$ contains the on-site energies, $\hat{H}_t$ describes
ligand-mediated hopping, and $\hat{H}_{\mathrm{SOC}}$ introduces SOC. The interplay of SOC with the broken inversion symmetry in the
asymmetric M$_1$–X–M$_2$ geometry generates the antisymmetric exchange responsible for
the DMI.

The on-site term corresponds to:
\begin{equation}
\hat{H}_0 = 
\sum_{im,\sigma}\varepsilon_{d,m}\,
d_{im\sigma}^\dagger d_{im\sigma}
+
\sum_{\alpha,\sigma}\varepsilon_{p,\alpha}\,
p_{\alpha\sigma}^\dagger p_{\alpha\sigma} +\sum_{\alpha,\sigma}\varepsilon_{d,\alpha}\,
d_{\alpha\sigma}^\dagger d_{\alpha\sigma} 
\end{equation}
where $\varepsilon_{d,m}$ and $\varepsilon_{p,\alpha}$ denote the on-site energies of the $m$-th M-$d$  and $\alpha$-th X-$p$ (or or X-$d$) orbitals. The operators
$d_{m\sigma}^\dagger$ ($d_{m\sigma}$) and $p_{\alpha\sigma}^\dagger$
($p_{\alpha\sigma}$) create (annihilate) an electron with spin $\sigma$ in the
corresponding orbitals. The ligand-mediated hopping between the magnetic and nonmagnetic sites is given by
\begin{equation}
\hat{H}_t = \sum_{i,m,\alpha,\sigma}
\left( H^{(iX)}_{m\alpha}\,
d_{im\sigma}^\dagger p_{\alpha\sigma} +H^{(iX)}_{m\alpha}\,
d_{im\sigma}^\dagger d_{\alpha\sigma}
+ \mathrm{h.c.} \right),
\end{equation}
while SOC contributes via
\begin{align}
\hat{H}_{\mathrm{SOC}} =
\lambda_X
\sum_{\alpha,\beta,\sigma,\sigma'}
(&
p_{\alpha\sigma}^\dagger\,
\langle p_\alpha | \mathbf{L} | p_\beta \rangle
\cdot
\langle \sigma | \mathbf{S} | \sigma' \rangle
p_{\beta\sigma'}
\\
&+
d_{\alpha\sigma}^\dagger\,
\langle d_\alpha | \mathbf{L} | d_\beta \rangle
\cdot
\langle \sigma | \mathbf{S} | \sigma' \rangle
d_{\beta\sigma'}
).
\end{align}
with analogous terms on the metal sites.

From the Wannier representation, we extract the orbital-resolved hopping matrix
elements for both Si and Pt ligands:
\begin{equation}
\begin{aligned}
t_{m,\alpha}^{(M_1\mathrm{Si})} &= 
\langle d_m^{(M_1)} | \hat{H} | p_\alpha^{(\mathrm{Si})} \rangle, \qquad
t_{n,\beta}^{(M_2\mathrm{Si})} =
\langle d_n^{(M_2)} | \hat{H} | p_\beta^{(\mathrm{Si})} \rangle,\\[4pt]
t_{m,\alpha}^{(M_1\mathrm{Pt})} &=
\langle d_m^{(M_1)} | \hat{H} | d_\alpha^{(\mathrm{Pt})} \rangle, \qquad\;
t_{n,\beta}^{(M_2\mathrm{Pt})} =
\langle d_n^{(M_2)} | \hat{H} | d_\beta^{(\mathrm{Pt})} \rangle,
\end{aligned}
\end{equation}
where $m,n,\alpha,\beta \in \{ xy, yz, xz, x^2-y^2, z^2 \}$ for $d$-type orbitals,
and $\alpha,\beta \in \{ x, y, z \}$ for $p$-type orbitals.

From the Wannier-based TB Hamiltonian, we extract all
orbital-resolved hopping amplitudes $t_{ij}$ between orbitals $i$ and $j$. For each hopping channel, the associated geometric direction is defined as
\begin{equation}
\vec{\mathbf R} = \mathbf a_{\mathrm{lat}} + (\mathbf r_i - \mathbf r_j),
\end{equation}
where $\mathbf a_{\mathrm{lat}}$ connects the unit cells of the two orbitals, and
$\mathbf r_i$, $\mathbf r_j$ denote their intracell coordinates.  The
corresponding hopping vector is then written as
\begin{equation}
\vec{\mathbf R}_{ij} = t_{ij}\,\vec{\mathbf R},
\end{equation}
which carries units of (eV$\,\cdot\,$\AA) and encodes both the magnitude and the
spatial orientation of the orbital couplings along the M$_1$–X–M$_2$ 
superexchange pathway, the microscopic origin of DMI.

Following Bl\"ugel~\cite{blugel}, the emergence of antisymmetric exchange is
governed by the noncollinearity of such hopping channels.  
For the M$_1$–X–M$_2$ trimer, the site-dependent orbital moments
$\vec{\theta}_i$ can be expressed as
\[
\vec{\theta}_i = \theta_i (\cos\phi\,\hat{\mathbf e}_x \pm
\sin\phi\,\hat{\mathbf e}_y),
\]
where $\phi$ is the canting angle defined in Fig.~\ref{tb1}(b).  
SOC and local inversion-symmetry breaking modify the orientation of the M$_i$–X
hybrid orbitals, and the resulting hopping vectors may be represented as
\[
\vec{\mathbf R}_{ij}=R_{ij}^{x}\hat{\mathbf e}_x
\pm R_{ij}^{y}\hat{\mathbf e}_y,
\]
with approximate magnitudes $|R_{ij}^{x}|\approx|\theta_i\cos\phi|$ and 
$|R_{ij}^{y}|\approx|\theta_i\sin\phi|$.
Once all hopping vectors are expressed in a common global coordinate frame, the
microscopic chirality of a given superexchange path is evaluated from the cross
product of the hopping channels involving the ligand,
\begin{equation}
d_{ij}^{\mathrm{TB}} \propto 
\mathbf R_{ik} \times \mathbf R_{kj},
\label{dmi-eq}
\end{equation}
where $i\in\mathrm{M}_1$, $j\in\mathrm{M}_2$, and $k$ labels the ligand orbital.
This quantity has units of (eV$^{2}$\AA$^{2}$) but is used only as a 
dimensionless chirality indicator: its sign and relative magnitude reveal
whether a particular M–X–M pathway supports finite antisymmetric exchange.  It
is directly analogous to the DMI term in Eq.~\ref{Hamiltomnian}, although it is
not a physical DMI coefficient.

The TB model is constructed from Wannier90 derived hopping parameters
\cite{pizzi2020wannier90}, with SOC strengths. The TB eigenvalues and eigenfunctions are fitted to the DFT band structure,
and excellent agreement is obtained for Pt$_{0.5}$Co$_{1}$Ti$_{0.5}$Si, as shown
in Fig.~S3 of the Supplementary Material~\cite{SM}. This Wannier-based TB analysis identifies two distinct antisymmetric exchange
paths: a weak Si-mediated channel and a dominant Pt-mediated interlayer channel.
Using Eq.~\ref{dmi-eq}, we compute the chirality indicators for the Co–Si–Co,
Co–Pt–Co, Ti–Si–Ti, and Ti–Pt–Ti paths (in both intralayer and interlayer
geometries) across all Pt/Co-substituted Ti$_2$Si compositions.  
The results in Table~\ref{tab2} show that nonzero entries identify
symmetry allowed chiral pathways, whereas the Ti--Si--Ti channel vanishes for
all compositions except Pt$_{0.5}$Ti$_{1.5}$Si. This composition behaves
differently because, in the absence of Co, only the weaker Ti-mediated pathways
are active, yielding smaller DMI signatures compared to the
Co-containing systems. The sign of $d_{ij}^{\mathrm{TB}}$ follows the
right-hand convention, with positive (negative) values indicating CCW (CW)
rotation around the $+\hat{z}$ ($-\hat{z}$) direction.
\\

\begin{table}[t]
\centering
\caption{
Tight-binding chirality indicators 
using Eq.~\ref{dmi-eq}
for the Co–Si–Co, Co–Pt–Co, Ti–Si–Ti, and Ti–Pt–Ti superexchange pathways in
Pt/Co–substituted Ti$_2$Si. 
These quantities have units of (eV$^{2}$\AA$^{2}$) but are reported in arbitrary 
units, since only their sign and relative magnitude determine whether a given 
M$_1$--X--M$_2$ pathway supports antisymmetric exchange.}

\begin{tabular}{ccccc}
\hline
\hline
\justifying
Material &Co-Si-Co&Co-Pt-Co&Ti-Si-Ti&Ti-Pt-Ti\\
\hline
Pt$_{0.5}$Ti$_{1.5}$Si  & 0.00 &0.00&-0.90&-2.08 \\
Pt$_{0.5}$Co$_{0.5}$Ti$_{1}$Si &0.89&-2.96&-0.002&0.00\\
Pt$_{0.5}$Co$_{1}$Ti$_{0.5}$Si  &4.38&-6.44&-0.00&0.00\\
CoTiSi                        &-1.68&0.00&0.00&0.00\\
PtTiSi &0.00&0.00&0.00&0.00\\
\hline
\hline
\end{tabular}
\label{tab2}
\end{table}

\begin{figure}
    \centering
 \includegraphics[width=1.04\linewidth]{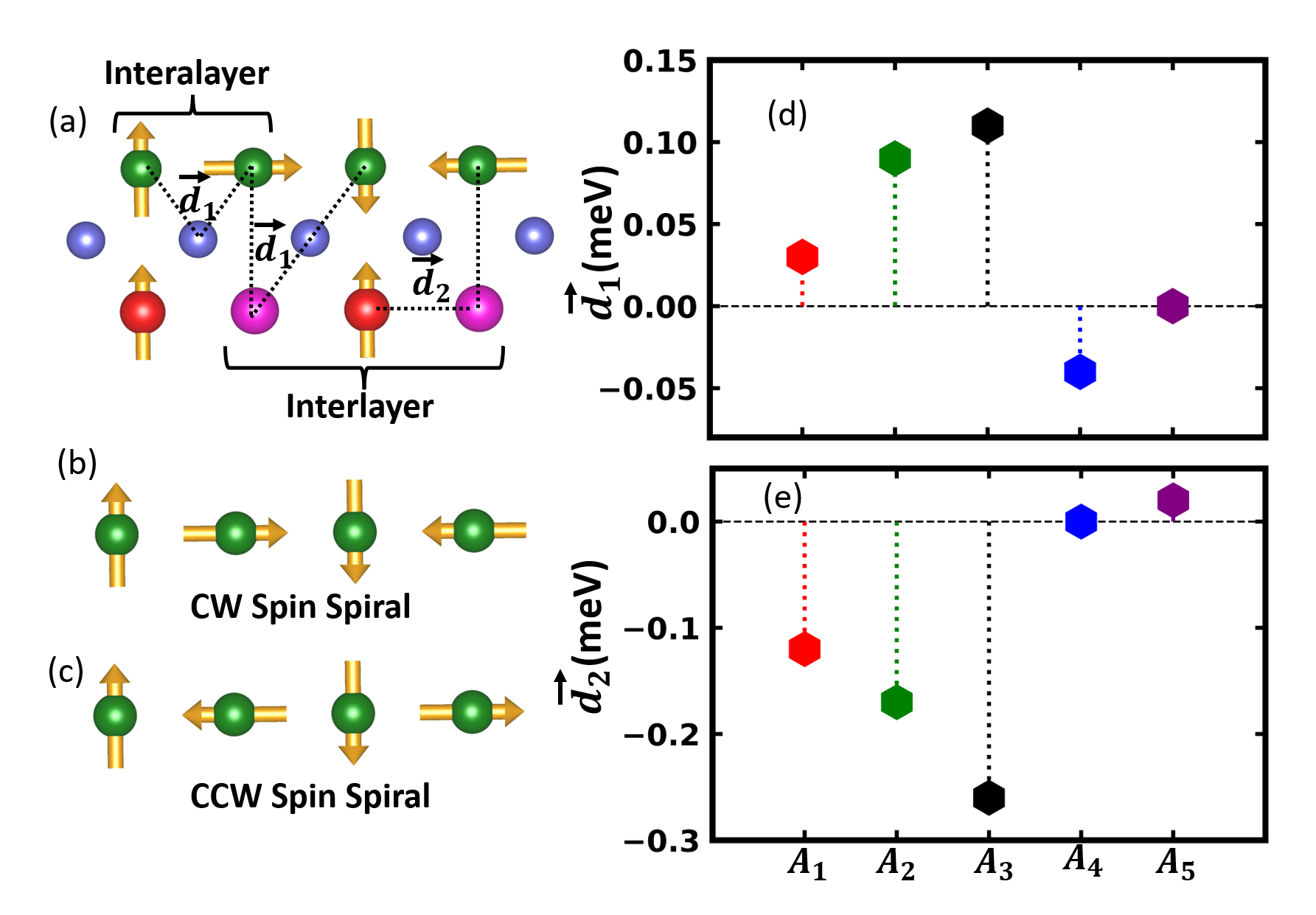}
 \caption{
(a) Schematic illustration of the two symmetry-allowed DMI pathways in doped 
Ti$_2$Si. The first pathway, $\vec{d}_{1}$, contains both intralayer and weak 
interlayer contributions transmitted primarily through Si and Pt, whereas the second 
pathway, $\vec{d}_{2}$, is a predominantly interlayer DMI mediated by Pt. 
(b) - (c) show the clockwise (CW) and Counterclockwise (CCW) spin spirals configurations. 
(d) and (e) Evolution of the calculated $d_{1}$ and $d_{2}$ values, respectively, 
for the configurations A$_1$ = Pt$_{0.5}$Ti$_{1.5}$Si, 
A$_2$ = Pt$_{0.5}$Co$_{0.5}$Ti$_{1}$Si, 
A$_3$ = Pt$_{0.5}$Co$_{1}$Ti$_{0.5}$Si, 
A$_4$ = CoTiSi, and 
A$_5$ = PtTiSi.
}

    \label{dm}
\end{figure}

\textit{Density functional results.} 
The TB model provides the microscopic chirality signatures, while the actual 
DMI vectors are obtained from a symmetry analysis of the real-space atomic 
geometry. Following Moriya’s symmetry rules~\cite{dmi,dmi2}, the DMI interaction 
between two magnetic atoms M$_1$ and M$_2$ located at sites $i$ and $j$ arises 
through the M$_1$--X--M$_2$ superexchange pathways mediated by the nonmagnetic atom(s) 
X, as illustrated in Fig.~\ref{tb1}. For each exchange path, the corresponding 
DMI vector takes the minimal symmetry-allowed form:

\begin{equation}
    \vec{d}_{ij} = d_{ij}\,(\hat{r}_{iX} \times \hat{r}_{jX}),
\end{equation}
where $\hat{r}_{iX}$ and $\hat{r}_{jX}$ are the unit vectors from the mediating 
atom X toward the magnetic sites $i$ and $j$. As illustrated in Fig.~\ref{dm}   (a), the present geometry contains two 
symmetry related M$_1$--X--M$_2$ pathways, giving rise to two independent DMI vectors: 
$\vec{d}_{1}$, which originates predominantly from the Si-mediated intralayer and interlayer
exchange, and $\vec{d}_{2}$, which is dominated by the Pt-mediated interlayer 
interaction.

The layer-resolved DMI constants are extracted using the spin-spiral method, where each component ($\vec{d}_1$ or $\vec{d}_2$) is obtained from the energy difference between clockwise (CW) and counterclockwise (CCW) spirals,
\[
d = \frac{E_{\mathrm{CW}} - E_{\mathrm{CCW}}}{m},
\]
with $m$ denoting the number of chiral magnetic bonds participating in the imposed spiral. The CW and CCW spin-spiral configurations used to extract the DMI constants are shown in Fig.~\ref{dm} (b) - (c). Since the structural models for the doped systems are already constructed using a $4\times2$ supercell, no additional supercell is required for the DMI calculations. Where $m=16$ for the Si mediated DMI vector $\vec{d}_1$, and $m=2\sqrt{2}$ for the Pt mediated interlayer DMI vector $\vec{d}_2$, as derived in Sec.~IV of the Supplementary Material~\cite{SM}. 

The resulting DMI vectors $\vec{d}_1$ and $\vec{d}_2$ are illustrated in Fig.~\ref{dm} (d) - (e), and their magnitudes are summarized in Table~\ref{tab1}. A clear hierarchy emerges: $\vec{d}_2$ is consistently and substantially larger than $\vec{d}_1$, reflecting the much stronger SOC on Pt and the more efficient chiral exchange it mediates across the spacer. This trend is fully consistent with our  TB analysis, which shows that the Co-Pt-Co exchange channel carries a much larger antisymmetric DMI exchange energy compared to the Ti-Si-Ti channel as shown in Table~\ref{tab2}. For Pt containing compositions such as Pt$_{0.5}$Ti$_{1.5}$Si, Pt$_{0.5}$Co$_{0.5}$Ti$_1$Si, and Pt$_{0.5}$Co$_{1}$Ti$_{0.5}$Si, the Pt assisted interlayer DMI dominates the total chiral interaction, in agreement with the TB prediction that Co–Pt-Co couplings exhibit the strongest SOC-induced asymmetry. In contrast, Si-mediated DMI remains weak: in CoTiSi, the absence of Pt leads to a small $\vec{d}_1$ due to Co-Si-Co consistent with TB results as shown in Table~\ref{tab2} and vanishing $\vec{d}_2$. Even in PtTiSi (where Co is absent), the Pt-mediated long-range pathway produces
a very small but finite DMI contribution, $d_{2}=-0.02$, while the Si-mediated
term $d_{1}$ essentially vanishes. The corresponding TB chirality indicators for
this composition appear nearly zero, reflecting the fact that both channels are
extremely weak and fall below the TB model's numerical resolution. Nevertheless,
the sign of the finite DFT-derived $d_{2}$ remains consistent with the general
TB trend that the Pt-mediated exchange is intrinsically stronger than the
Si-mediated one. Taken together, these results demonstrate that Pt driven interlayer DMI is the primary source of chirality in doped Ti$_2$Si, while Si-mediated intralayer and interlayer contributions are intrinsically weak.

In the micromagnetic limit, the discrete lattice spins $\vec{S}_{i}$ are replaced by a continuous magnetization field $\vec{m}(\mathbf{r})$. Under this transformation, the DMI contribution arising from the third term of the Hamiltonian in Eq.~\ref{Hamiltomnian} can be expressed as
\begin{equation}
E_{\mathrm{DMI}}
= \int d\mathbf{r}\epsilon_{\mathrm{DMI}}
= \int d\mathbf{r}\sum_{\mu} D_{\mu}\big[\vec{m}\times\partial_{\mu}\vec{m}\big],
\end{equation}
where $\epsilon_{\mathrm{DMI}}=\sum_{\mu} D_{\mu},(\vec{m}\times\partial_{\mu}\vec{m})$ is the micromagnetic DMI energy density, $\partial_{\mu}$ denotes the spatial derivative along direction $\mu$, and $D_{\mu}$ is the corresponding micromagnetic DMI vector. For interfacial systems, this expression reduces to the widely used form~\cite{iwasaki2013current,sampaio2013nucleation,fert2013skyrmions,yang2023first}
\begin{equation}
E_{\mathrm{DMI}}
= \int d\mathbf{r}
D\left[
\hat{\mathbf{m}}\cdot(\nabla\times\hat{\mathbf{m}})
-(\hat{\mathbf{m}}\cdot\nabla)\hat{\mathbf{m}}
\right]\cdot\hat{\mathbf{z}},
\label{dmi}
\end{equation}
where $D$ is the effective interfacial micromagnetic DMI constant.

The micromagnetic coefficients $D_{1}$ and $D_{2}$ are connected to their corresponding atomistic DMI vectors $\vec{d_{1}}$ and $\vec{d_{2}}$ (derived in Supplementary~\cite{SM} Sec.~V; see Eqs.~S25 and S29) through
\begin{equation}
D_{1} = \frac{d_{1}N}{at}, \label{D1}
\end{equation}
\begin{equation}
D_{2} = \frac{d_{2}N}{at}, \label{D2}
\end{equation}
where $N$ is the number of magnetic atoms per unit cell, $a$ is the in-plane lattice parameter, and $t$ is the monolayer thickness. The extracted micromagnetic DMI values are listed in Table~\ref{tab1}.

Among all compositions, Pt$_{0.5}$Co$_{1}$Ti$_{0.5}$Si exhibits the strongest 
chiral interaction, with $D_{1}=0.21$~mJ/m$^{2}$ and $D_{2}=-0.50$~mJ/m$^{2}$. 
The opposite signs of $D_{1}$ and $D_{2}$ reflect their distinct rotational 
preferences: $D_{1}$ favors a CCW spin spiral, whereas $D_{2}$ favors a CW 
spiral, consistent with the conventions used in Eqs.~\ref{D1} and \ref{D2}. Pt$_{0.5}$Co$_{0.5}$Ti$_{1}$Si likewise displays finite chirality, with 
$D_{1}=0.17$~mJ/m$^{2}$ (CCW) and $D_{2}=-0.33$~mJ/m$^{2}$ (CW). These results 
show that both intralayer (Si-mediated) and interlayer (Si- and Pt-mediated) 
exchange pathways contribute to the total DMI, with the Pt-assisted path 
dominant in magnitude. The rotation senses extracted from $D_{1}$ and $D_{2}$ 
are also fully consistent with the TB chirality analysis as shown in Table~\ref{tab2}. Overall, the atomistic and micromagnetic results demonstrate that Pt-driven 
interlayer DMI is the primary source of chiral exchange in doped Ti$_2$Si, 
whereas the Si-mediated intralayer contribution remains intrinsically weak. 
This establishes Ti$_2$Si monolayers with controlled Pt and Co substitution as 
a robust platform for engineering sizable and tunable DMI, enabling the 
stabilization of chiral spin textures in centrosymmetry broken 2D 
TMS.


\section{Conclusion}
In summary, we have combined density-functional theory, micromagnetic
analysis, and Wannier-based tight-binding modeling to uncover the
conditions under which monolayer Ti$_2$Si can host chiral magnetic
interactions. The pristine material is a ferromagnetic metal whose
exchange coupling and magnetic anisotropy can be tuned by an external
electric field, but the underlying inversion symmetry prevents the
formation of DMI. By introducing controlled Pt and Co substitution at
Ti sites, we break the horizontal mirror symmetry of the Ti–Si–Ti
trilayer and simultaneously enhance both magnetic exchange (via Co) and
SOC (via Pt). This chemical asymmetry generates a robust
DMI, dominated by a Pt-mediated interlayer superexchange channel, as 
confirmed by first-principles spiral calculations and independently by a
TB chirality analysis. The strongest DMI occurs in the
Pt$_{0.5}$CoTi$_{0.5}$Si composition, where both intralayer favours CCW spin spiral rotation and interlayer
paths consistently favor a CW rotation sense. Overall, our results demonstrate that substitutional doping provides an 
effective, experimentally accessible route to engineer sizable and tunable 
DMI in doped 2D silicides. Ti$_2$Si-based monolayers thus constitute 
a promising platform for realizing electrically controllable chiral spin 
textures and for developing next-generation 2D spintronic devices.

\twocolumngrid
\bibliography{reference}
\bibliographystyle{apsrev}

\end{document}